%% file: Blind_De-anonymization_Attacks_using_Social_Networks.tex
\documentclass[sigconf]{acmart}
\usepackage{booktabs} 
\usepackage{graphicx}
\usepackage{balance}  
\usepackage{enumerate}
\usepackage{epstopdf}
\usepackage{amsmath,amsthm}
\usepackage{subfigure}
\usepackage{tabu}
\usepackage{algorithm}
\usepackage{algorithmic}
\usepackage[algo2e]{algorithm2e}

\usepackage{etoolbox}
\makeatletter
\patchcmd{\maketitle}{\@copyrightspace}{}{}{}
\makeatother
\usepackage{epsfig}
\usepackage{color}
\usepackage{bm}

\newcommand{\wpes}{\textcolor{black}}
\setcopyright{acmcopyright}

\acmDOI{10.1145/3139550.3139562}

\acmISBN{978-1-4503-5175-1/17/10}

\acmConference{WPES'17}{October 30, 2017}{Dallas, TX, USA.}
\acmYear{2017}
\copyrightyear{2017}

\acmPrice{15.00}

\begin{document}
\title{Blind De-anonymization Attacks using Social Networks }
\fancyhead{} 

\author{Wei-Han Lee}
\affiliation{%
  \institution{Princeton University}
}
\email{weihanl@princeton.edu}

\author{Changchang Liu}
\affiliation{%
  \institution{Princeton University}
}
\email{cl12@princeton.edu}

\author{Shouling Ji}
\affiliation{%
  \institution{Zhejiang University, China and Georgia Tech, USA}
}
\email{sji@gatech.edu}

\author{Prateek Mittal}
\affiliation{%
  \institution{Princeton University}
}
\email{pmittal@princeton.edu}

\author{Ruby B. Lee}
\affiliation{%
  \institution{Princeton University}
}
\email{rblee@princeton.edu}

\begin{abstract}
It is important to study the risks of publishing privacy-sensitive data. Even if sensitive identities (e.g., name, social security number) were removed and advanced data perturbation techniques were applied, several de-anonymization attacks have been proposed to re-identify individuals. However, existing attacks have some limitations: 1) they are limited in de-anonymization accuracy; 2) they require prior seed knowledge and suffer from the imprecision of such seed information.

We propose a novel structure-based de-anonymization attack, which does not require the attacker to have prior information (e.g., seeds). Our attack is based on two key insights: using multi-hop neighborhood information, and optimizing the process of de-anonymization by exploiting enhanced machine learning techniques. The experimental results demonstrate that our method is robust to data perturbations and significantly outperforms the state-of-the-art de-anonymization techniques by up to $10\times$ improvement.
\end{abstract}

%
%
%
%
%
\keywords{De-anonymization; Graph Anonymity; Machine Learning;}

\settopmatter{printacmref=false}
\maketitle

\input{intro}
\input{method}

\input{experiments}

\vspace{-0.5em}
\input{conclusion}
\vspace{-1.5em}
\bibliographystyle{ACM-Reference-Format}
\bibliography{bib} 

\end{document}

%% file: intro.tex
\section{Introduction}
Privacy-sensitive data (social relationships, mobility traces, medical records, etc.) are increasingly becoming public to facilitate data-mining researchers and applications. To protect users' privacy, data anonymization techniques have been the focus of extensive investigations \cite{hay2007anonymizing, liu2008SIGMOD, Pedarsani2011SIGKDD}.

Most privacy-sensitive data are closely related to individual behavior, and thus contain rich structural/graph-theoretic characteristics. For example, social networks can be modeled as graphs in a straightforward manner. Mobility traces can also be modeled as graph topologies \cite{Srivatsa2012CCS}. However, even equipped with sophisticated anonymization techniques \cite{dwork2008differential,liu2016dependence,liu2016linkmirage}, the privacy of structural data still suffers from de-anonymization attacks assuming that the adversaries have access to rich auxiliary information (also called background information) from other channels \cite{backstrom2007WWW, Narayanan2008SP, Narayanan2009SP,hay2008VLDB, liu2008SIGMOD, Srivatsa2012CCS,ji2014CCS,nilizadeh2014CCS,li2016enhanced,chiasserini2015impact,kazemi2015growing, lee2017quantification, lee2017how}. 

Today, many individuals have accounts in various social networks such as Facebook, Twitter, Google+, Myspace and Flickr. Based on the inherent cross-site correlations, Narayanan et al. \cite{Narayanan2009SP} effectively de-anonymized a Twitter dataset by utilizing a Flickr dataset as auxiliary information. Furthermore, Nilizadeh et al. \cite{nilizadeh2014CCS} exploited the community structure of a graph to de-anonymize social networks. Other public datasets may also contain individual behavior information. For instance, Srivatsa et al. \cite{Srivatsa2012CCS} proposed to de-anonymize a set of location traces based on a social network. They demonstrated that a contact graph identifying meetings between anonymized users in the location traces can be structurally correlated with the corresponding social network graph.

However, previous work on de-anonymization attacks  {have} several limitations: 1) most previous works~\cite{Narayanan2008SP,Narayanan2009SP,nilizadeh2014CCS}  {rely} on a seed-identification process. To obtain the useful seeds, they assume that the attacker possesses detailed information about a small number of members of the target network. They also assume that the attacker can determine if these members are also present in his auxiliary network (e.g., by matching user names and other contextual information). Furthermore, these methods may suffer from the imprecision of the adversary's background knowledge (misidentified seeds); 2) existing seed-free de-anonymization techniques \cite{ji2014CCS, pedarsani2013Allerton} have limited accuracy because they only  {utilize} limited structural information of the data. In this paper, we aim to solve these problems by proposing a novel blind (i.e., seed-free) de-anonymization technique and exploring fine-grained structure information of graph topologies. Overall, we make the following contributions:
\begin{figure*}[!t]
\centering
\epsfig{file=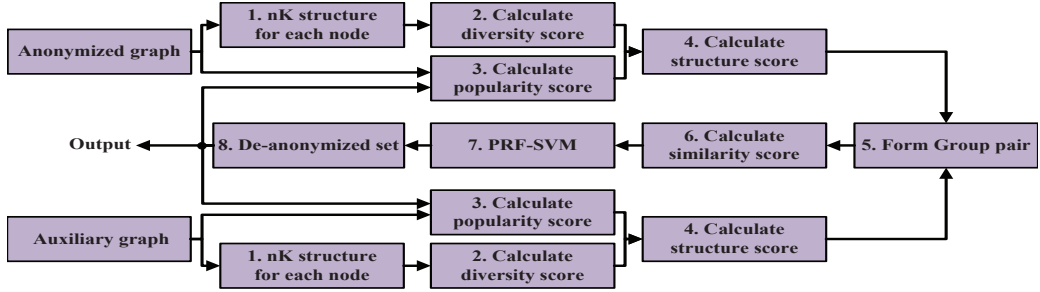, width=5.4in, height=1.5in}
\caption{{Mechanism for our blind de-anonymization attack. 
}}
\label{mechanism}
\end{figure*}
\begin{enumerate}[$\bullet$]
\item{We present a novel de-anonymization technique, which does not require adversaries to have any prior information (e.g., seeds). \wpes{In our method, 1) we propose the \emph{$\bf{nK}$-series} to incorporate multi-hop neighbors' information in graph structures as novel features in our de-anonymization attack;} 2) we jointly optimize the matching for users between the anonymized graph and the auxiliary graph by leveraging a machine learning technique: pseudo relevance feedback support vector machine (PRF-SVM).}

\item{\wpes{We show that our method is practical and effective: our attack is robust to data perturbations and has significant de-anonymization advantages over existing approaches with up to $10\times$ improvement. Our method demonstrates that structural data can be effectively de-anonymized even without any seed information.}
}
\end{enumerate}

%% file: method.tex
\section{Blind De-anonymization Attacks}\label{attack}
Previous works on structure-based de-anonymization  {do} not fully utilize fine-grained graph-theoretic information. For instance, only one-hop neighbors have been utilized in \cite{Narayanan2008SP}, and very limited structural information has been leveraged in \cite{ji2014CCS}. Also, most existing work  {rely} on assumptions that the adversary has prior information or some ground truth (e.g., the seed information in \cite{Narayanan2009SP}). However, seed-based de-anonymization attacks have some issues in practice: 1) the seed identification process usually requires heavy computational complexity \cite{Narayanan2009SP}; 2) misidentified seeds may seriously decrease the de-anonymization capability. 

We propose a general, blind (seed-free) de-anonymization attack. Figure \ref{mechanism} outlines our method, which consists of the following key steps on each of the anonymized graph, $G_a$, and the auxiliary graph, $G_u$. Our proposed $\bf{nK}$-series aims to capture fine-grained structural information of each node, and the PRF-SVM aims to jointly de-anonymize the nodes. By exploring richer and finer-grained structural information of the graphs, our method can achieve better de-anonymization performance without requiring the adversaries to have any prior information (e.g., seeds). 

\emph{\bf Step 1:} We first explore multi-hop neighbor information by proposing our new \emph{$\bf{nK}$-series} structural features for each node.

\emph{$\bf{nK}$-series:} Inspired by the idea of the $\bf{dK}$-series~\cite{Mahadevan2006SIGCOMM} for characterizing structural statistics of a graph, we propose the $\bf{nK}$-series to describe structural features of each node in a fine-grained manner, by incorporating the structural information of its multi-hop neighbors. $\bf{nK0}$ represents the degree of the node, i.e., the number of its neighbors. 
$\bf{nK1}$ captures the degree histogram of its neighbors and $\bf{nK2}$ captures the degree histogram of its 2-hop neighbors. Here, we focus our research on $\bf{nK0}$, $\bf{nK1}$, and $\bf{nK2}$ to construct the $\bf{nK}$ structural features of each node $a$ as ${\bf{v}}(a)=\left[{\bf{nK0}}(a), {\bf{nK1}}(a), {\bf{nK2}}(a)\right]^{T}$.


\emph{\bf Step 2:} Based on the $\bf{nK}$ structural features, we calculate the \emph{diversity score} for each node $a$, which measures the richness of the structural characteristics of this node and is defined as $DS(a)=\frac{\sum_{i} \tilde{v_i}(a)\log {\tilde{v_i}(a)}}{\log(\dim({\tilde{\bf{v}}(a)}))}$.
where $\tilde{\bf{v}}(a)$ denotes the normalized structural feature vector of $a$, i.e., $\tilde{\bf{v}}(a) = \frac{{\bf{v}(a)}}{\|{\bf{v}(a)}\|_2}$. $\dim({\tilde{\bf{v}}(a)})$ denotes the dimension for ${\tilde{\bf{v}}(a)}$. Here, $\sum_{i} \tilde{v_i}(a)\log {\tilde{v_i}(a)}$ is actually similar to \emph{entropy} in information theory \cite{cover:JW12}, which evaluates the amount of information stored in $\tilde{v_i}(a)$, and $\log(\dim({\tilde{\bf{v}}(a)}))$ is just for normalizing the diversity score so that $DS(a)\in [0,1]$. A higher diversity score means that this node has more distinguishable structural characteristics.

Next, we start de-anonymizing the anonymized data in an iterative manner.

\emph{\bf Step 3:} For each round, we calculate the \emph{popularity score} for each node, which evaluates its relationships with the set of de-anonymized nodes in the previous round (the set of de-anonymized nodes is empty in the initial round). We denote $\mathcal{N}^t$ as the set of nodes that have been de-anonymized after the $t$-th iteration, where $\mathcal{N}^t$ is an empty set for the first round. We define the popularity score of node $a$, $PS(a)$, as the Jaccard similarity \cite{real1996probabilistic} between the set of neighbors $N(a)$ for each node $a$ and $\mathcal{N}^t$ as: 
\begin{equation}
PS(a) = J(\mathcal{N}^t,N(a))={|\mathcal{N}^t\cap N(a)|}/{|\mathcal{N}^t\cup N(a)|}
\end{equation}
where $J(A,B)$ is the Jaccard similarity, $J(A,B) = \frac{|A\cap B|}{|A\cup B|}$, and $PS(a)\in[0,1]$. A higher popularity score represents a closer relationship between this anonymized node and those previously de-anonymized nodes. In addition to the diversity score, the popularity score can also be leveraged to evaluate the structural characteristics of each anonymized node.

\emph{\bf Step 4:} Subsequently, we compute the \emph{structure score} ($SS$) for each node $a$ as $SS(a) = DS(a)+c\cdot PS(a)$. 
 where $c$ is a pre-defined parameter to balance the diversity score and the popularity score. 

\emph{\bf Step 5:} \wpes{Next, we group the nodes in the anonymized graph and the auxiliary graph according to their structure scores. Our grouping process works as follows: for the $t$-th iteration, we select $N_{group}$ nodes with higher $SS$ from the anonymized graph and the auxiliary graph to form the group pair $C_a^t$ (for the anonymized graph) and $C_u^t$ (for the auxiliary graph). Note that for the first iteration, we select those nodes with higher $DS$ (since $PS=0$ for the initial round).}

\emph{\bf Step 6:} For each group pair, we rank each potential pair of nodes according to the similarities between their $\bf{nK}$ structural features.

For each node $a$ in $C_a^t$ and node $b$ in $C_u^t$, we evaluate the similarity between their $\bf{nK}$ structural features by computing the cosine similarity \cite{garcia2006cosim} between ${\bf{v}}(a)$ and ${\bf{v}}(b)$ as $\mathrm{Sim}(a,b)=\frac{\left<{\bf{v}}(a), {\bf{v}}(b)\right>}{\|{\bf{v}}(a)\|_2\|{\bf{v}}(b)\|_2}$. Larger cosine similarity score means two nodes are more similar. Furthermore, to emphasize the differences between node pairs and thus to improve the node matching performance, we can transform the above similarity linearly as
\begin{equation}\label{eq_sim}
S(a,b)=\max\limits_{d\in C_u^t}(\mathrm{Sim}(a,d))-\frac{\max\limits_{d\in C_u^t}(\mathrm{Sim}(a,d))-\mathrm{Sim}(a,b)}{{var}(\mathrm{Sim}(a,:))}
\end{equation}
where $\mathrm{Sim}(a,:)$ is a vector consisting of $\mathrm{Sim}(a,d)$ for $d \in C_u^t $ and ${var}(\cdot)$ is the variance of a vector.

\emph{\bf Step 7:} Next, we leverage machine learning techniques: pseudo relevance feedback with support vector machine (PRF-SVM), to re-rank these potential pairs of nodes. 

Specifically, we view this node-matching process from a classification perspective, i.e., we aim to classify all the possible pairs of nodes as two categories: \emph{matched} or \emph{unmatched}. For each SVM iteration, we select the top $N_{train}$ node pairs with the highest similarity scores and the bottom $N_{train}$ node pairs with the lowest similarity scores as the training data, labeling them as \emph{matched} and \emph{unmatched}, respectively.

With these training node pairs, we apply SVM to classify the remaining node pairs.
The SVM method would result in a classification hyperplane. Based on this hyperplane, each possible node pair would be given a value $dis(a,b)$ derived according to its distance from the hyperplane. We define a confidence score $\mathrm{SVM}(a,b)$ for each potential node pair $(a,b)$, which is linearly normalized as $\mathrm{SVM}(a,b)=\frac{|dis(a,b)|-d_{min}}{d_{max}-d_{min}}$, where $d_{max}, d_{min}$ represent the maximum and minimum distance from the hyperplane computed over all the remaining node pairs.  The updated similarity score $\hat{S}(a,b)$ is obtained by integrating the original similarity score $S(a,b)$ in Eq.~\ref{eq_sim} with the confidence score $\mathrm{SVM}(a,b)$ as
\begin{equation}\label{lee2}
\hat{S}(a,b)=S(a,b)\cdot \mathrm{SVM}(a,b)^\alpha
\end{equation}
where $\alpha$ is a parameter that emphasizes the importance of the confidence score SVM$(a,b)$ in Eq. \ref{lee2}. A new ranking list is thus generated based on these updated similarity scores, $\hat{S}(a,b)$. This process of classification and re-ranking can be conducted iteratively until a stable classification result is obtained.

\emph{\bf Step 8:} Finally, we extract the matched pairs of nodes based on the classification result of PRF-SVM in \emph{Step 7} and then update the set of de-anonymized nodes $\mathcal{N}^t$. We iteratively repeat \emph{Step 3}-\emph{Step 7} until we cannot de-anonymize any more nodes.

Note that although our method is seed-free, it can be directly generalized to incorporate seed knowledge if the adversary has such prior information. Given a set of known seeds, these seeds could be considered as the matched result in the first group of our algorithm, and the iteration for finding more matched nodes can be implemented consequently as shown in Figure \ref{mechanism}.

%

%% file: experiments.tex
\section{Experimental Analysis} \label{sec:experiment}
In this section, we compare our attack with the state-of-the-art de-anonymization techniques~\cite{ji2014CCS,nilizadeh2014CCS}, to show the significant advantage of our approach (up to $10\times$ improvement in de-anonymization accuracy). For fair comparison, we use the default parameters in the code these authors provided or the optimal parameters they utilized in their papers. We experiment on the collaboration dataset, the Twitter dataset and the Gowalla dataset (discussed below) for fair comparison with the method of Ji et al. and the method of Nilizadeh et al. since these are also the datasets they utilized \cite{ji2014CCS,nilizadeh2014CCS}.

\subsection{Datasets and General Settings}\label{sec:dataset}
The Collaboration dataset \cite{newman2001PNAS} is a network of co-authorships between scientists who have posted preprints on the Condensed Matter E-Print Archive, which consists of {\it 36,458} users and {\it 171,735} edges. The Twitter dataset \cite{nilizadeh2014CCS} captures the connections between users who mentioned each other at least once between March 24th, 2012 and April 25th, 2012, and contains two different graphs named Twitter (small) with {\it 9,745} users and {\it 50,164} edges, and Twitter (large) with {\it 90,331} users and {\it 358,422} edges. 

The Gowalla dataset consists of a social graph and a mobility trace dataset \cite{pham2013SIGMOD}. The social graph contains {\it 196,591} users with {\it 950,327} edges. The mobility trace consists of {\it 6.44}M checkins generated by these users. To better evaluate the performance of our method, we leverage the techniques in \cite{pham2013SIGMOD} to construct four graphs from the mobility trace dataset with different recalls and precisions, denoted by M1, M2, M3, and M4. All the four mobility trace graphs contain {\it 196,591} users, and the corresponding number of edges are {\emph{659,186}},~~~{\emph{829,375}},~~~{\emph{ 919,671}},~~~{\emph{1,070,790}}, respectively.

To evaluate the performance of our de-anonymization attack, we consider a popular perturbation method of Hay et al. \cite{hay2007anonymizing}, which applies a sequence of $r$ random edge deletions followed by $r$ random edge insertions (a similar perturbation process has been utilized for the de-anonymization attacks in \cite{nilizadeh2014CCS}). Here, we define $noise$ (perturbations) as the extent of edge modification, i.e., the ratio of altered edges $r$ to the total number of edges $M$, i.e., $noise=\frac{r}{M}$. Note that we add the same amount of $noise$ to the original graph of the {Collaboration, Twitter} datasets to obtain the anonymized graph and the auxiliary graph, respectively. For Gowalla mobility trace, we utilize its social network structure for de-anonymization attacks. \wpes{Furthermore, we vary the system parameters of our method and set $c=2, N_{group}=1000, N_{train}=1250, \alpha=1$ for achieving the best performance in our experiments.}

We utilize \emph{Accuracy} to evaluate the de-anonymization performance. Accuracy is the ratio of the correctly de-anonymized nodes out of all the overlapped nodes between the anonymized graph and the auxiliary graph, i.e., $\emph{Accuracy}=\frac{N_{cor}}{|V_a\cap V_u|}$, where $N_{cor}$ is the number of correctly de-anonymized nodes and $V_a, V_u$ represent the sets of nodes in the anonymized and auxiliary graph, respectively.


\begin{figure*}[!t]
\centering
\subfigure[{Collaboration}]{
\label{condense_ji}
\includegraphics[width=1.8in,height=0.9in]{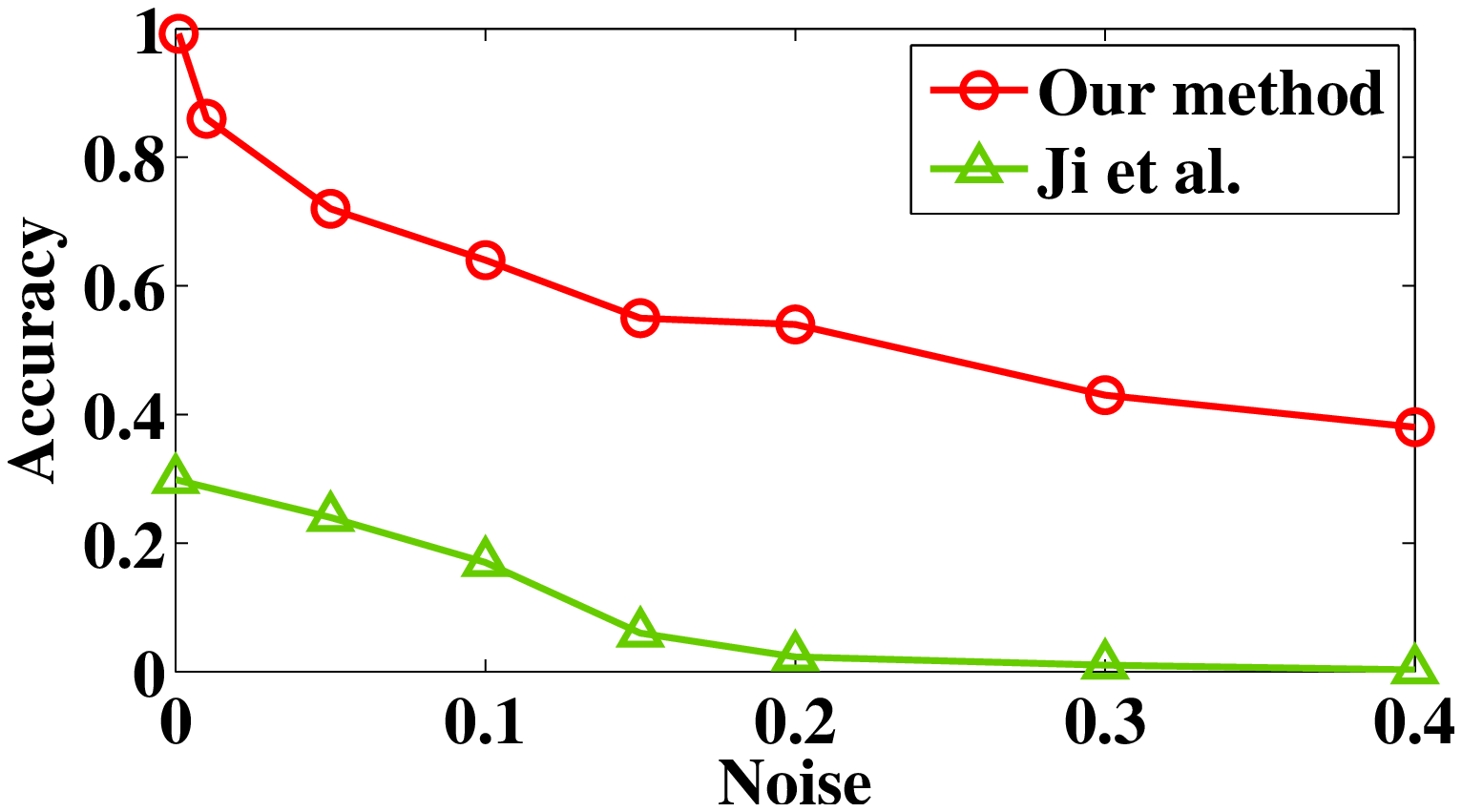}}
\hspace{-1.8em}
\subfigure[{Twitter: small}]{
\label{twitter1_ji}
\includegraphics[width=1.8in,height=0.9in]{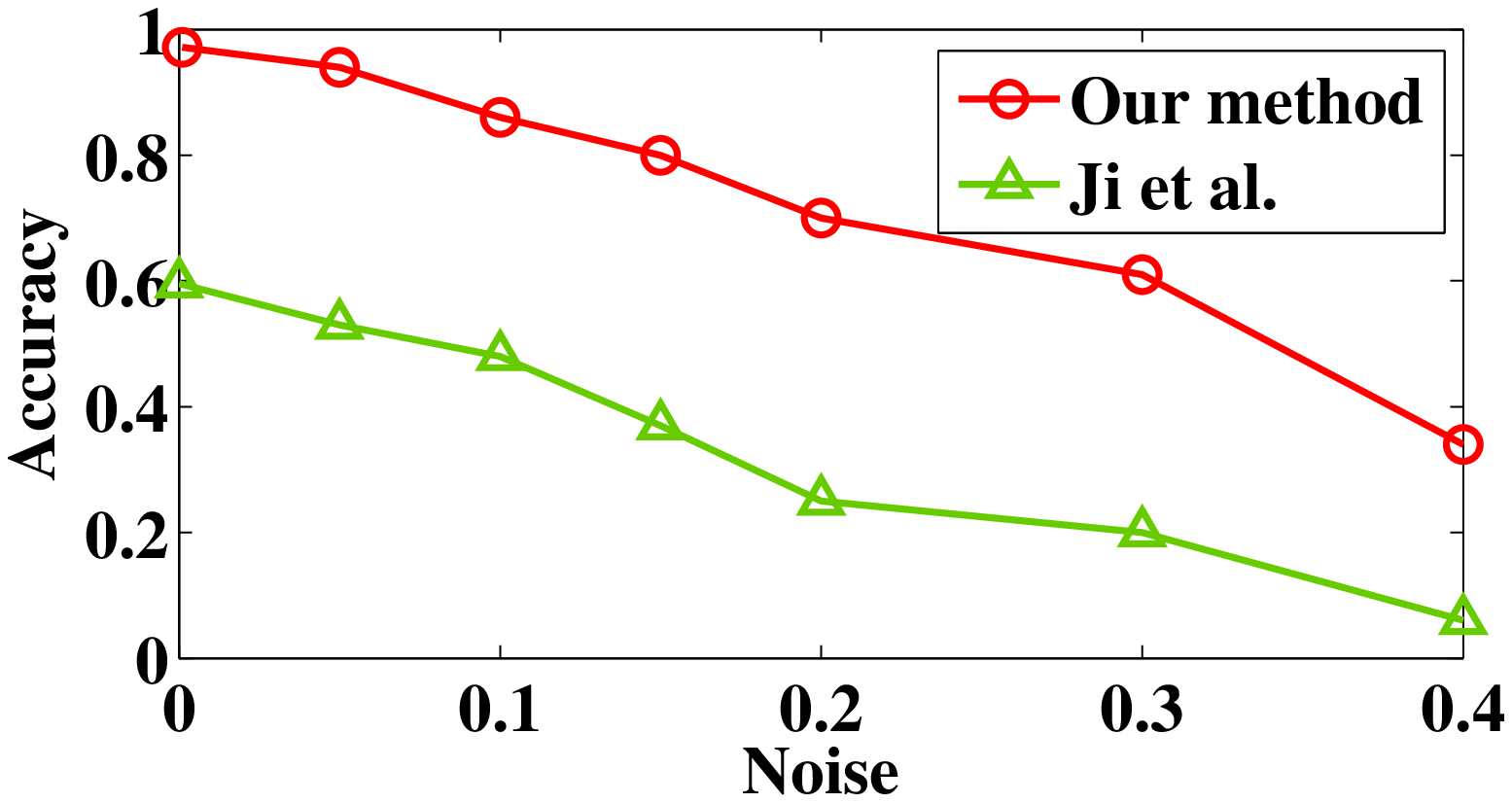}}
\hspace{-1.8em}
\subfigure[{Twitter: large}]{
\label{twitter2_ji}
\includegraphics[width=1.8in,height=0.9in]{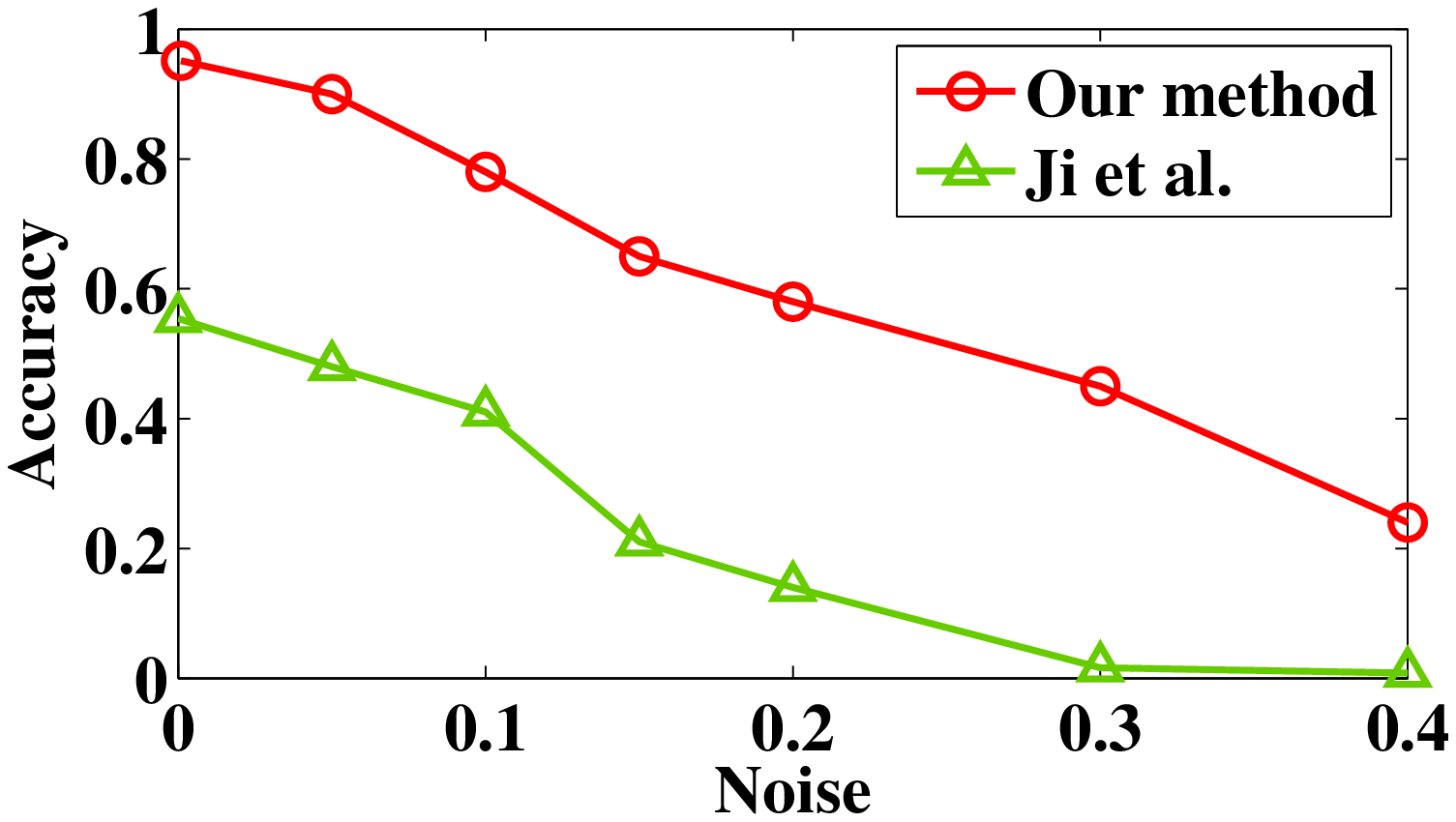}}
\hspace{-1.8em}
\subfigure[{Gowalla}]{
\label{gowalla}
\includegraphics[width=1.8in,height=0.9in]{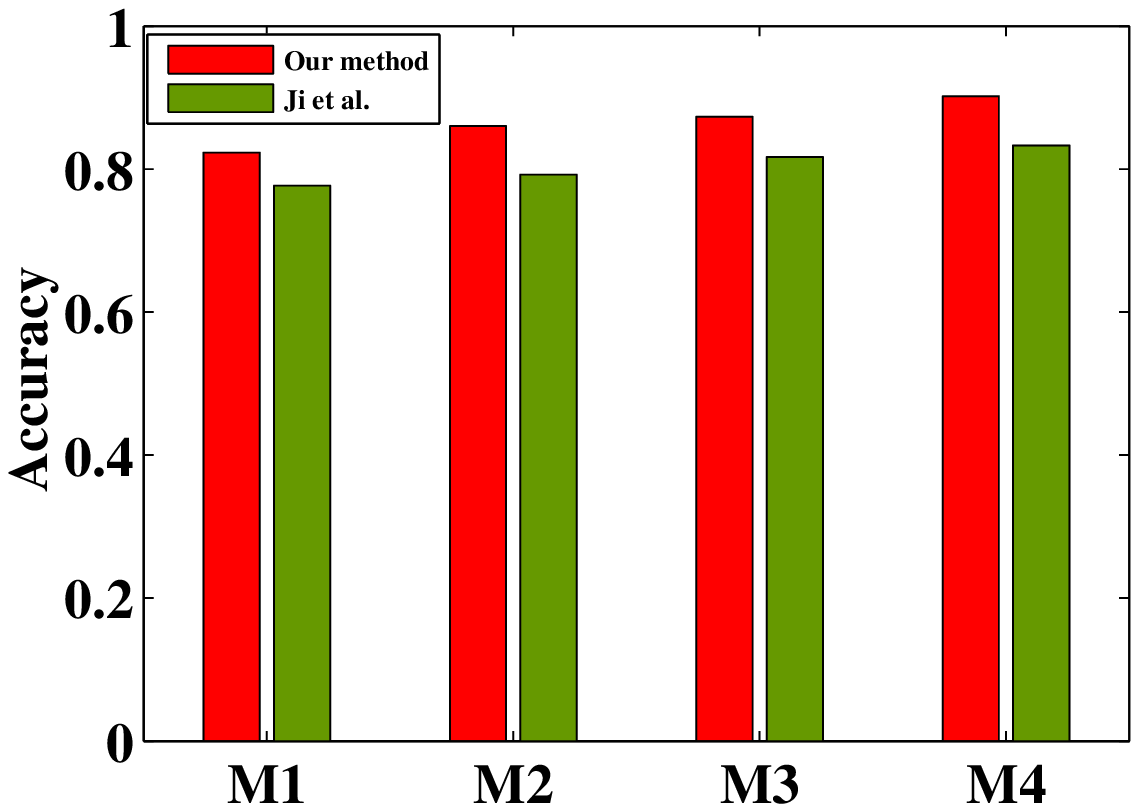}}
\vspace{-1.6em}
\caption{{The comparison of our approach with the method of Ji et al.~\cite{ji2014CCS}. 
}}
\label{twitter_ji} 
\end{figure*}

\begin{figure*}[!t]
\centering
\subfigure[{Collaboration}]{
\label{condense_mismatched}
\includegraphics[width=2in,height=0.9in]{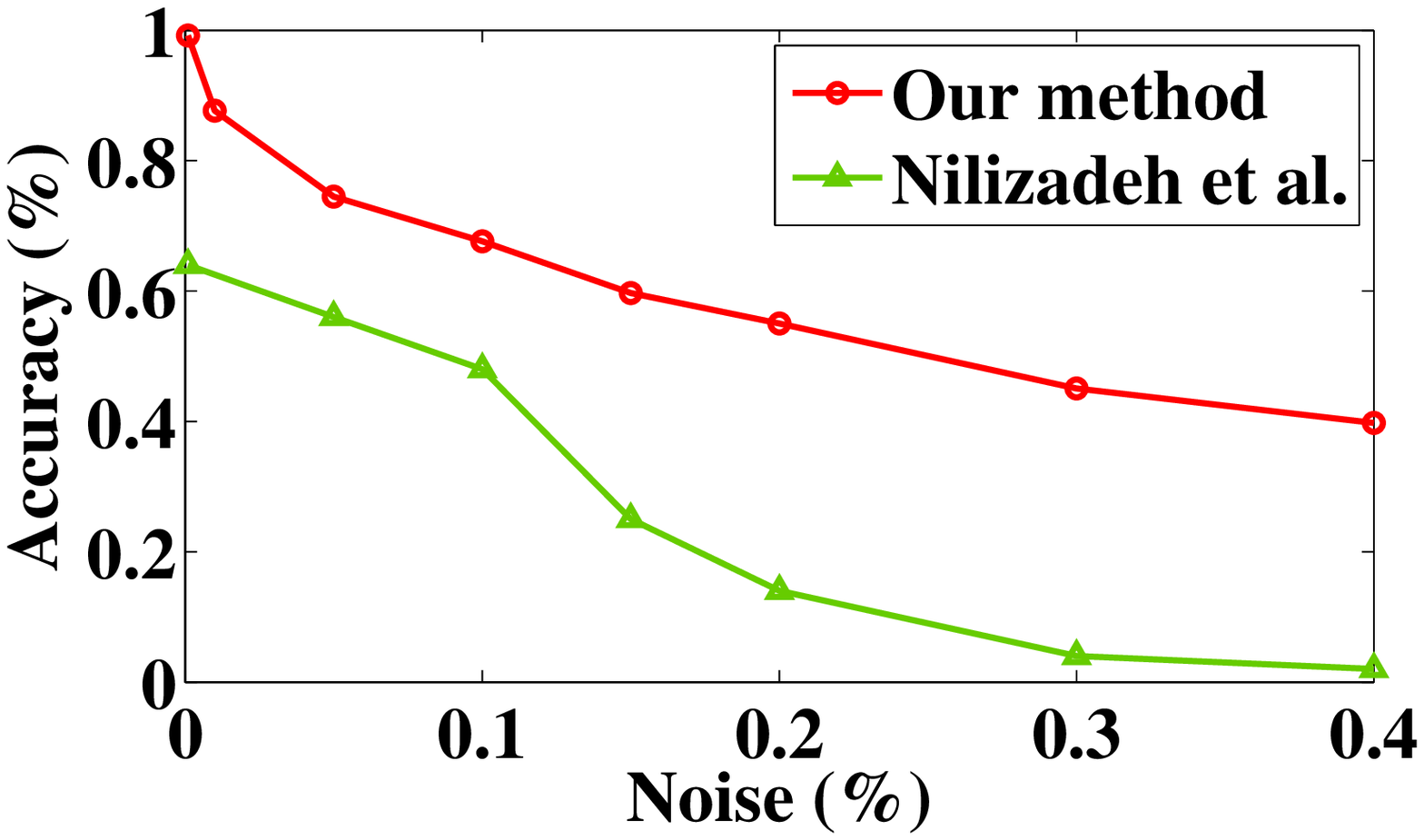}}
\hspace{-1em}
\subfigure[{Twitter: small}]{
\label{twitter1_mismatched}
\includegraphics[width=2in,height=0.9in]{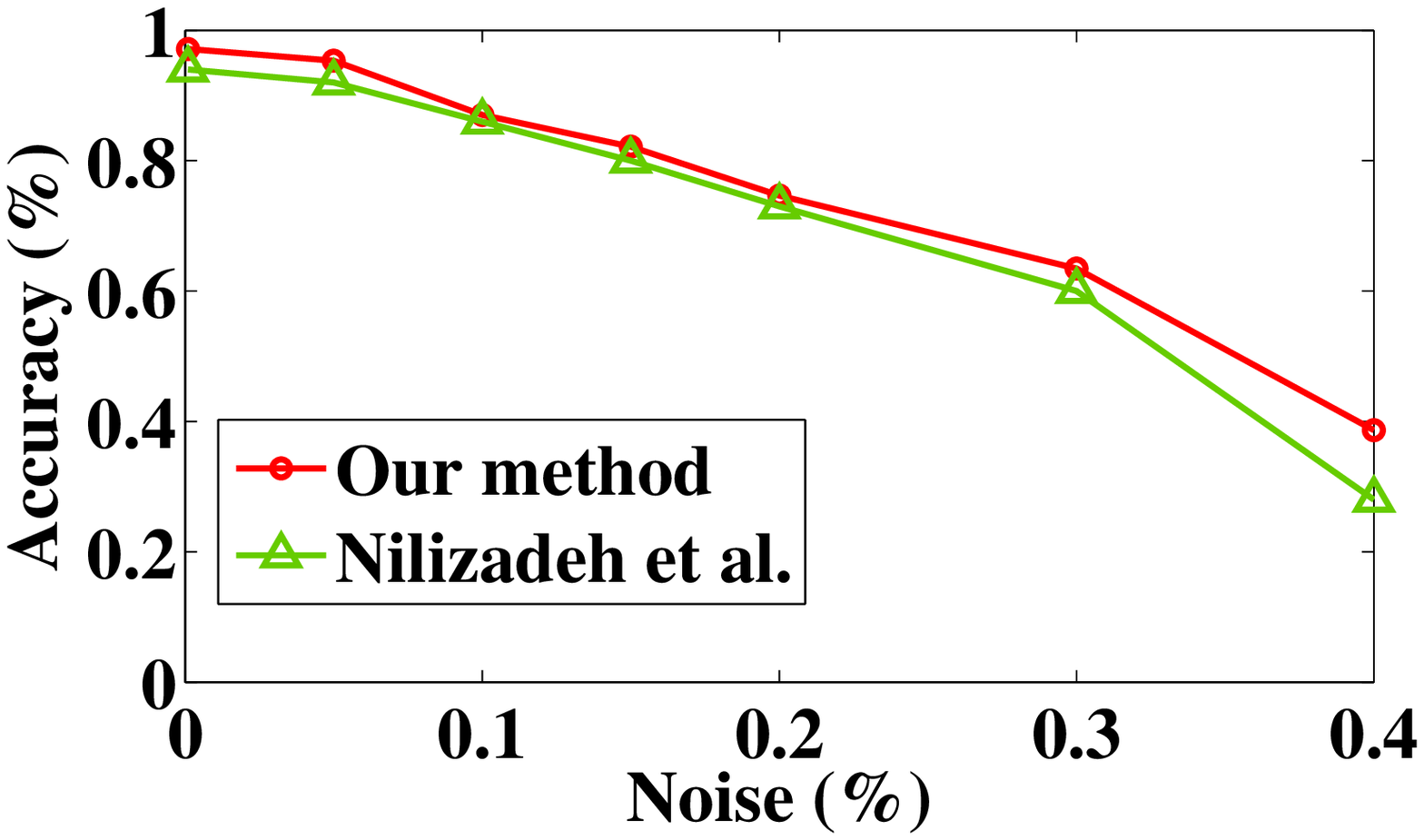}}
\hspace{-1em}
\subfigure[{Twitter: large}]{
\label{twitter2_mismatched}
\includegraphics[width=2in,height=0.9in]{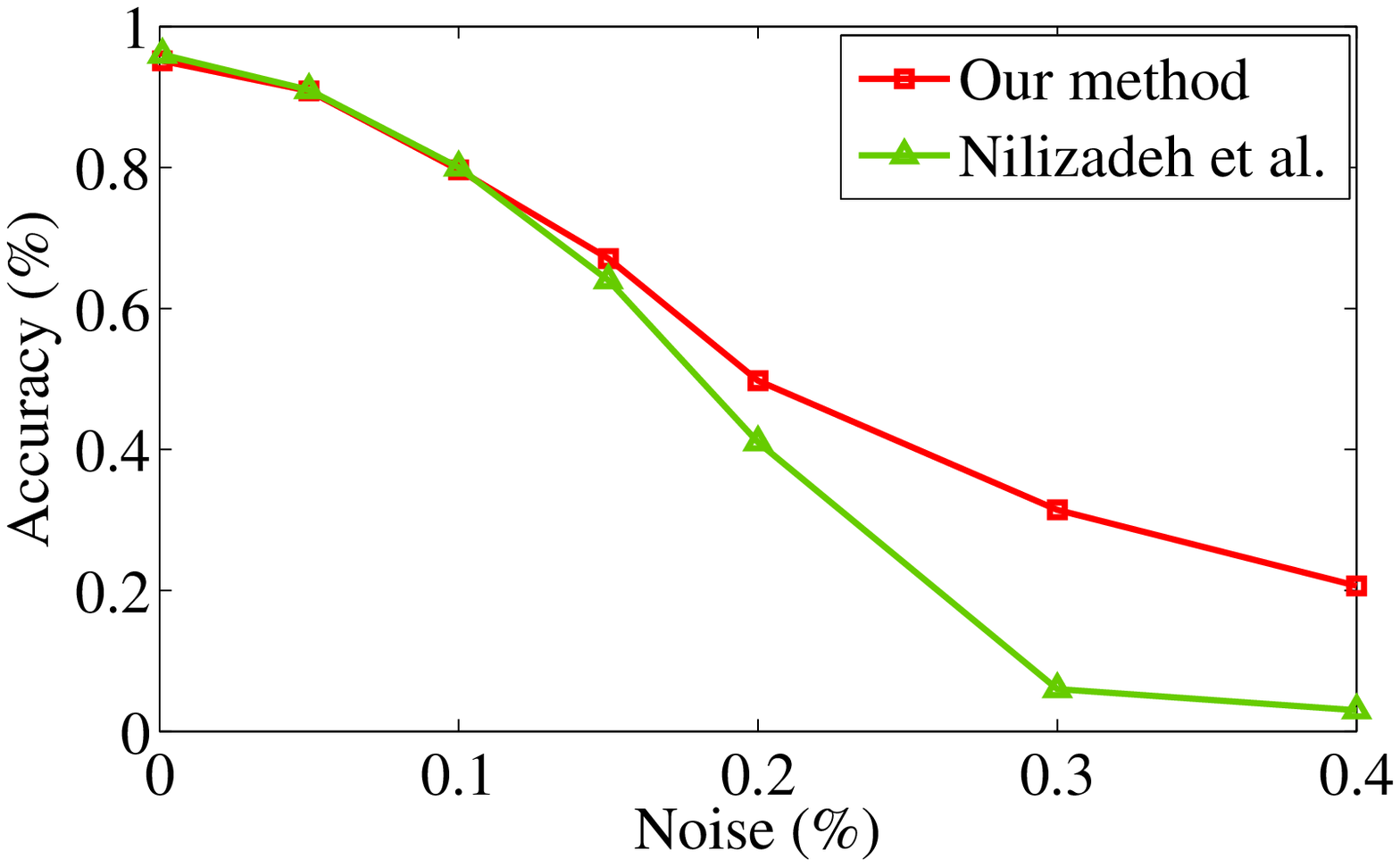}}
\vspace{-1.6em}
\caption{{The comparison of our approach with the method of Nilizadeh et al.~\cite{nilizadeh2014CCS}.
}}
\label{mismatched} 
\end{figure*}

\subsection{Comparison with Ji et al. \protect\cite{ji2014CCS}}\label{sec:com_ji}
Ji et al. \cite{ji2014CCS} proposed a cold-start optimization-based de-anonymization attack. Although they utilized four structural attributes for each node: degree, 1-hop neighborhood, top-K reference distance and sampling closeness centrality, these attributes only represent coarse-grained structure information of the graphs. 

We compare our approach with the method of Ji et al. in the Collaboration dataset, the Twitter dataset and the Gowalla dataset in Figure \ref{twitter_ji}. We can see that our approach has much higher accuracy than their method: we can achieve up to $10\times$ improvement for collaboration dataset, and about $6\times$ improvement for two Twitter graphs. Furthermore, we utilize the Gowalla social dataset to de-anonymize the Gowalla mobility trace dataset, in order to compare with the method of Ji et al. for fairness (they experimented on this data in \cite{ji2014CCS}). In Figure \ref{gowalla}, the de-anonymization results of our method for Gowalla mobility trace datasets (M1)(M2)(M3)(M4) are $81.3\%$, $84.8\%$, $85.3\%$ and $89.1\%$, respectively. By utilizing finer-grained and richer structural information, our method also outperforms the method of Ji et al. for the Gowalla dataset.

\subsection{Comparison with Nilizadeh et al. \protect\cite{nilizadeh2014CCS}}
Nilizadeh et al. \cite{nilizadeh2014CCS} leveraged community detection techniques to partition the networks into separate components. Then, they applied existing network alignment methods to the nodes inside the communities for more seed knowledge. However, their method has the following limitations: 1) it requires prior knowledge (seeds) to boot up their attack, which is a strong assumption and may suffer from misidentified seeds; 2) their performance may be influenced by the inconsistency problem of community detection methods~\cite{xie2013overlapping}. We experiment on the collaboration dataset and the two Twitter graphs for fair comparison with the method of Nilizadeh et al. since these are also the data they used in \cite{nilizadeh2014CCS}.

Figure \ref{condense_mismatched} compares our method with the approach of Nilizadeh et al. on the collaboration dataset. Our method can de-anonymize much more authors and is also more stable to data perturbations. \wpes{For $noise=0.4$, our method significantly outperforms the method of Nilizadeth et al. by more than $10\times$ for de-anonymization accuracy.}

Figure \ref{twitter1_mismatched} and \ref{twitter2_mismatched} compare our method with the method of Nilizadeh et al. on the Twitter datasets. Our method is more robust to noise, and has higher accuracy especially when the noise is high. For $noise=0.4$, we have almost $1.25\times$ improvement for Twitter (small) dataset, and $9\times$ improvement for Twitter (large) dataset.

%% file: conclusion.tex
\section{Conclusion}
We presented a novel blind (seed-free) de-anonymization method by utilizing the $\bf{nK}$-series that we define to capture fine-grained structure features, and proposing a new variant of the SVM machine learning technique called PRF-SVM to do concurrent matching of the nodes between the anonymized graph and the auxiliary graph. Experimental results demonstrate the significant advantages (up to $10\times$ improvement in de-anonymization accuracy) of our method over the state-of-the-art de-anonymization attacks.

%% file: Blind_De-anonymization_Attacks_using_Social_Networks.bbl

\begin{thebibliography}{00}


\ifx \showCODEN    \undefined \def \showCODEN     #1{\unskip}     \fi
\ifx \showDOI      \undefined \def \showDOI       #1{#1}\fi
\ifx \showISBNx    \undefined \def \showISBNx     #1{\unskip}     \fi
\ifx \showISBNxiii \undefined \def \showISBNxiii  #1{\unskip}     \fi
\ifx \showISSN     \undefined \def \showISSN      #1{\unskip}     \fi
\ifx \showLCCN     \undefined \def \showLCCN      #1{\unskip}     \fi
\ifx \shownote     \undefined \def \shownote      #1{#1}          \fi
\ifx \showarticletitle \undefined \def \showarticletitle #1{#1}   \fi
\ifx \showURL      \undefined \def \showURL       {\relax}        \fi
\providecommand\bibfield[2]{#2}
\providecommand\bibinfo[2]{#2}
\providecommand\natexlab[1]{#1}
\providecommand\showeprint[2][]{arXiv:#2}

\bibitem[\protect\citeauthoryear{Backstrom, Dwork, and Kleinberg}{Backstrom
  et~al\mbox{.}}{2007}]%
        {backstrom2007WWW}
\bibfield{author}{\bibinfo{person}{Lars Backstrom}, \bibinfo{person}{Cynthia
  Dwork}, {and} \bibinfo{person}{Jon Kleinberg}.}
  \bibinfo{year}{2007}\natexlab{}.
\newblock \showarticletitle{Wherefore art thou r3579x?: anonymized social
  networks, hidden patterns, and structural steganography}. In
  \bibinfo{booktitle}{{\em WWW}}.
\newblock


\bibitem[\protect\citeauthoryear{Chiasserini, Garetto, and
  Leonardi}{Chiasserini et~al\mbox{.}}{2015}]%
        {chiasserini2015impact}
\bibfield{author}{\bibinfo{person}{Carla-Fabiana Chiasserini},
  \bibinfo{person}{Michele Garetto}, {and} \bibinfo{person}{Emilio Leonardi}.}
  \bibinfo{year}{2015}\natexlab{}.
\newblock \showarticletitle{Impact of Clustering on the Performance of Network
  De-anonymization}. In \bibinfo{booktitle}{{\em COSN}}.
\newblock


\bibitem[\protect\citeauthoryear{Cover and Thomas}{Cover and Thomas}{2012}]%
        {cover:JW12}
\bibfield{author}{\bibinfo{person}{Thomas~M Cover} {and} \bibinfo{person}{Joy~A
  Thomas}.} \bibinfo{year}{2012}\natexlab{}.
\newblock \bibinfo{booktitle}{{\em Elements of information theory}}.
\newblock \bibinfo{publisher}{John Wiley \& Sons}.
\newblock


\bibitem[\protect\citeauthoryear{Dwork}{Dwork}{2008}]%
        {dwork2008differential}
\bibfield{author}{\bibinfo{person}{Cynthia Dwork}.}
  \bibinfo{year}{2008}\natexlab{}.
\newblock \showarticletitle{Differential privacy: A survey of results}. In
  \bibinfo{booktitle}{{\em International Conference on Theory and Applications
  of Models of Computation}}. Springer, \bibinfo{pages}{1--19}.
\newblock


\bibitem[\protect\citeauthoryear{Garcia}{Garcia}{2006}]%
        {garcia2006cosim}
\bibfield{author}{\bibinfo{person}{E Garcia}.} \bibinfo{year}{2006}\natexlab{}.
\newblock \showarticletitle{Cosine similarity and term weight tutorial}.
\newblock \bibinfo{journal}{{\em Information retrieval intelligence\/}}
  (\bibinfo{year}{2006}).
\newblock


\bibitem[\protect\citeauthoryear{Hay, Miklau, Jensen, Towsley, and Weis}{Hay
  et~al\mbox{.}}{2008}]%
        {hay2008VLDB}
\bibfield{author}{\bibinfo{person}{Michael Hay}, \bibinfo{person}{Gerome
  Miklau}, \bibinfo{person}{David Jensen}, \bibinfo{person}{Don Towsley}, {and}
  \bibinfo{person}{Philipp Weis}.} \bibinfo{year}{2008}\natexlab{}.
\newblock \showarticletitle{Resisting structural re-identification in
  anonymized social networks}. In \bibinfo{booktitle}{{\em VLDB}}.
\newblock


\bibitem[\protect\citeauthoryear{Hay, Miklau, Jensen, Weis, and Srivastava}{Hay
  et~al\mbox{.}}{2007}]%
        {hay2007anonymizing}
\bibfield{author}{\bibinfo{person}{Michael Hay}, \bibinfo{person}{Gerome
  Miklau}, \bibinfo{person}{David Jensen}, \bibinfo{person}{Philipp Weis},
  {and} \bibinfo{person}{Siddharth Srivastava}.}
  \bibinfo{year}{2007}\natexlab{}.
\newblock \showarticletitle{Anonymizing social networks}.
\newblock \bibinfo{journal}{{\em Computer Science Department Faculty
  Publication Series\/}} (\bibinfo{year}{2007}).
\newblock


\bibitem[\protect\citeauthoryear{Ji, Li, Srivatsa, and Beyah}{Ji
  et~al\mbox{.}}{2014}]%
        {ji2014CCS}
\bibfield{author}{\bibinfo{person}{Shouling Ji}, \bibinfo{person}{Weiqing Li},
  \bibinfo{person}{Mudhakar Srivatsa}, {and} \bibinfo{person}{Raheem Beyah}.}
  \bibinfo{year}{2014}\natexlab{}.
\newblock \showarticletitle{Structural data de-anonymization: Quantification,
  practice, and implications}. In \bibinfo{booktitle}{{\em ACM CCS}}.
\newblock


\bibitem[\protect\citeauthoryear{Kazemi, Hassani, and Grossglauser}{Kazemi
  et~al\mbox{.}}{2015}]%
        {kazemi2015growing}
\bibfield{author}{\bibinfo{person}{Ehsan Kazemi}, \bibinfo{person}{S~Hamed
  Hassani}, {and} \bibinfo{person}{Matthias Grossglauser}.}
  \bibinfo{year}{2015}\natexlab{}.
\newblock \showarticletitle{Growing a graph matching from a handful of seeds}.
  In \bibinfo{booktitle}{{\em VLDB}}.
\newblock


\bibitem[\protect\citeauthoryear{Lee, Liu, Ji, Mittal, and Lee}{Lee
  et~al\mbox{.}}{2017a}]%
        {lee2017how}
\bibfield{author}{\bibinfo{person}{Wei-Han Lee}, \bibinfo{person}{Changchang
  Liu}, \bibinfo{person}{Shouling Ji}, \bibinfo{person}{Prateek Mittal}, {and}
  \bibinfo{person}{Ruby Lee}.} \bibinfo{year}{2017}\natexlab{a}.
\newblock \showarticletitle{How to Quantify Graph De-anonymization Risks}.
\newblock In \bibinfo{booktitle}{{\em Information Systems Security and
  Privacy}}. \bibinfo{publisher}{Springer}.
\newblock


\bibitem[\protect\citeauthoryear{Lee, Liu, Ji, Mittal, and Lee}{Lee
  et~al\mbox{.}}{2017b}]%
        {lee2017quantification}
\bibfield{author}{\bibinfo{person}{Wei-Han Lee}, \bibinfo{person}{Changchang
  Liu}, \bibinfo{person}{Shouling Ji}, \bibinfo{person}{Prateek Mittal}, {and}
  \bibinfo{person}{Ruby~B Lee}.} \bibinfo{year}{2017}\natexlab{b}.
\newblock \showarticletitle{Quantification of De-anonymization Risks in Social
  Networks}. In \bibinfo{booktitle}{{\em Information Systems Security and
  Privacy}}. IEEE.
\newblock


\bibitem[\protect\citeauthoryear{Li, Zhang, He, Cheng, Liu, and Sun}{Li
  et~al\mbox{.}}{2016}]%
        {li2016enhanced}
\bibfield{author}{\bibinfo{person}{Hong Li}, \bibinfo{person}{Cheng Zhang},
  \bibinfo{person}{Yunhua He}, \bibinfo{person}{Xiuzhen Cheng},
  \bibinfo{person}{Yan Liu}, {and} \bibinfo{person}{Limin Sun}.}
  \bibinfo{year}{2016}\natexlab{}.
\newblock \showarticletitle{An enhanced structure-based de-anonymization of
  online social networks}. In \bibinfo{booktitle}{{\em WASA}}.
\newblock


\bibitem[\protect\citeauthoryear{Liu, Chakraborty, and Mittal}{Liu
  et~al\mbox{.}}{2016}]%
        {liu2016dependence}
\bibfield{author}{\bibinfo{person}{Changchang Liu}, \bibinfo{person}{Supriyo
  Chakraborty}, {and} \bibinfo{person}{Prateek Mittal}.}
  \bibinfo{year}{2016}\natexlab{}.
\newblock \showarticletitle{Dependence Makes You Vulnberable: Differential
  Privacy Under Dependent Tuples.}. In \bibinfo{booktitle}{{\em NDSS}}.
\newblock


\bibitem[\protect\citeauthoryear{Liu and Mittal}{Liu and Mittal}{2016}]%
        {liu2016linkmirage}
\bibfield{author}{\bibinfo{person}{Changchang Liu} {and}
  \bibinfo{person}{Prateek Mittal}.} \bibinfo{year}{2016}\natexlab{}.
\newblock \showarticletitle{LinkMirage: Enabling Privacy-preserving Analytics
  on Social Relationships.}. In \bibinfo{booktitle}{{\em NDSS}}.
\newblock


\bibitem[\protect\citeauthoryear{Liu and Terzi}{Liu and Terzi}{2008}]%
        {liu2008SIGMOD}
\bibfield{author}{\bibinfo{person}{Kun Liu} {and} \bibinfo{person}{Evimaria
  Terzi}.} \bibinfo{year}{2008}\natexlab{}.
\newblock \showarticletitle{Towards identity anonymization on graphs}. In
  \bibinfo{booktitle}{{\em SIGMOD}}.
\newblock


\bibitem[\protect\citeauthoryear{Mahadevan, Krioukov, Fall, and
  Vahdat}{Mahadevan et~al\mbox{.}}{[n. d.]}]%
        {Mahadevan2006SIGCOMM}
\bibfield{author}{\bibinfo{person}{Priya Mahadevan}, \bibinfo{person}{Dmitri
  Krioukov}, \bibinfo{person}{Kevin Fall}, {and} \bibinfo{person}{Amin
  Vahdat}.} \bibinfo{year}{[n. d.]}\natexlab{}.
\newblock \showarticletitle{Systematic topology analysis and generation using
  degree correlations}. In \bibinfo{booktitle}{{\em SIGCOMM}}. ACM.
\newblock


\bibitem[\protect\citeauthoryear{Narayanan and Shmatikov}{Narayanan and
  Shmatikov}{2008}]%
        {Narayanan2008SP}
\bibfield{author}{\bibinfo{person}{Arvind Narayanan} {and}
  \bibinfo{person}{Vitaly Shmatikov}.} \bibinfo{year}{2008}\natexlab{}.
\newblock \showarticletitle{Robust de-anonymization of large sparse datasets}.
  In \bibinfo{booktitle}{{\em IEEE S\&P}}.
\newblock


\bibitem[\protect\citeauthoryear{Narayanan and Shmatikov}{Narayanan and
  Shmatikov}{2009}]%
        {Narayanan2009SP}
\bibfield{author}{\bibinfo{person}{Arvind Narayanan} {and}
  \bibinfo{person}{Vitaly Shmatikov}.} \bibinfo{year}{2009}\natexlab{}.
\newblock \showarticletitle{De-anonymizing social networks}. In
  \bibinfo{booktitle}{{\em IEEE S\&P}}.
\newblock


\bibitem[\protect\citeauthoryear{Newman}{Newman}{2001}]%
        {newman2001PNAS}
\bibfield{author}{\bibinfo{person}{Mark~EJ Newman}.}
  \bibinfo{year}{2001}\natexlab{}.
\newblock \showarticletitle{The structure of scientific collaboration
  networks}.
\newblock \bibinfo{journal}{{\em Proceedings of the National Academy of
  Sciences\/}} (\bibinfo{year}{2001}).
\newblock


\bibitem[\protect\citeauthoryear{Nilizadeh, Kapadia, and Ahn}{Nilizadeh
  et~al\mbox{.}}{2014}]%
        {nilizadeh2014CCS}
\bibfield{author}{\bibinfo{person}{Shirin Nilizadeh}, \bibinfo{person}{Apu
  Kapadia}, {and} \bibinfo{person}{Yong-Yeol Ahn}.}
  \bibinfo{year}{2014}\natexlab{}.
\newblock \showarticletitle{Community-enhanced de-anonymization of online
  social networks}. In \bibinfo{booktitle}{{\em ACM CCS}}.
\newblock


\bibitem[\protect\citeauthoryear{Pedarsani, Figueiredo, and
  Grossglauser}{Pedarsani et~al\mbox{.}}{2013}]%
        {pedarsani2013Allerton}
\bibfield{author}{\bibinfo{person}{Pedram Pedarsani}, \bibinfo{person}{Daniel~R
  Figueiredo}, {and} \bibinfo{person}{Matthias Grossglauser}.}
  \bibinfo{year}{2013}\natexlab{}.
\newblock \showarticletitle{A bayesian method for matching two similar graphs
  without seeds}. In \bibinfo{booktitle}{{\em Allerton Conference on
  Communication, Control, and Computing}}.
\newblock


\bibitem[\protect\citeauthoryear{Pedarsani and Grossglauser}{Pedarsani and
  Grossglauser}{2011}]%
        {Pedarsani2011SIGKDD}
\bibfield{author}{\bibinfo{person}{Pedram Pedarsani} {and}
  \bibinfo{person}{Matthias Grossglauser}.} \bibinfo{year}{2011}\natexlab{}.
\newblock \showarticletitle{On the privacy of anonymized networks}. In
  \bibinfo{booktitle}{{\em SIGMOD}}.
\newblock


\bibitem[\protect\citeauthoryear{Pham, Shahabi, and Liu}{Pham
  et~al\mbox{.}}{2013}]%
        {pham2013SIGMOD}
\bibfield{author}{\bibinfo{person}{Huy Pham}, \bibinfo{person}{Cyrus Shahabi},
  {and} \bibinfo{person}{Yan Liu}.} \bibinfo{year}{2013}\natexlab{}.
\newblock \showarticletitle{EBM: an entropy-based model to infer social
  strength from spatiotemporal data}. In \bibinfo{booktitle}{{\em SIGMOD}}.
\newblock


\bibitem[\protect\citeauthoryear{Real and Vargas}{Real and Vargas}{1996}]%
        {real1996probabilistic}
\bibfield{author}{\bibinfo{person}{Raimundo Real} {and} \bibinfo{person}{Juan~M
  Vargas}.} \bibinfo{year}{1996}\natexlab{}.
\newblock \showarticletitle{The probabilistic basis of Jaccard's index of
  similarity}.
\newblock \bibinfo{journal}{{\em Systematic biology\/}} (\bibinfo{year}{1996}).
\newblock


\bibitem[\protect\citeauthoryear{Srivatsa and Hicks}{Srivatsa and
  Hicks}{2012}]%
        {Srivatsa2012CCS}
\bibfield{author}{\bibinfo{person}{Mudhakar Srivatsa} {and}
  \bibinfo{person}{Mike Hicks}.} \bibinfo{year}{2012}\natexlab{}.
\newblock \showarticletitle{Deanonymizing mobility traces: Using social network
  as a side-channel}. In \bibinfo{booktitle}{{\em ACM CCS}}.
\newblock


\bibitem[\protect\citeauthoryear{Xie, Kelley, and Szymanski}{Xie
  et~al\mbox{.}}{2013}]%
        {xie2013overlapping}
\bibfield{author}{\bibinfo{person}{Jierui Xie}, \bibinfo{person}{Stephen
  Kelley}, {and} \bibinfo{person}{Boleslaw~K Szymanski}.}
  \bibinfo{year}{2013}\natexlab{}.
\newblock \showarticletitle{Overlapping community detection in networks: The
  state-of-the-art and comparative study}.
\newblock \bibinfo{journal}{{\em Acm computing surveys (csur)\/}}
  \bibinfo{volume}{45}, \bibinfo{number}{4} (\bibinfo{year}{2013}),
  \bibinfo{pages}{43}.
\newblock


\end{thebibliography}
